\documentclass[Universe,article,accept,moreauthors,dvi2pdf,12pt,a4paper]{mdpi}

\lastpage{185}
\doinum{10.3390/universe1020173}
\pubvolume{1}
\pubyear{2015}
\history{Received: 21 May 2015 / Accepted: 29 July 2015  / Published: xx}
\pdfoutput=1

\usepackage[latin1]{inputenc}

\usepackage{dcolumn}
\usepackage{bm}
\usepackage{ifpdf}
\usepackage{hyperref}
\usepackage{dcolumn}
\usepackage{bm}
\usepackage[spanish,english]{babel}
\usepackage{amsfonts}
\usepackage{amssymb}
\usepackage{graphicx}
\usepackage[latin1]{inputenc}
\usepackage{soul}
\usepackage{booktabs}
\usepackage{multirow}

\newcommand \be{\begin{equation}}
\newcommand \en{\end{equation}}
\newcommand \bea{\begin{eqnarray}}
\newcommand \ena{\end{eqnarray}}

\Title{Nonsingular Black Holes in $f(R)$ Theories}

\Author{Gonzalo J. Olmo $^{1,2,}$*, D. Rubiera-Garcia $^{3}$}

\address{%
$^{1}$ Departamento de F\'{i}sica Te\'{o}rica and IFIC, Centro Mixto Universidad de
Valencia - CSIC. Universidad de Valencia, Burjassot-46100, Valencia, Spain\\
$^{2}$ Departamento de F\'isica, Universidade Federal da
Para\'\i ba, 58051-900 Jo\~ao Pessoa, Para\'\i ba, Brazil \\
$^{3}$ Center for Field Theory and Particle Physics and Department of Physics, Fudan University, \linebreak 220 Handan Road, 200433 Shanghai, China; E-Mail: drubiera@fudan.edu.cn\vspace{-12pt}}

\corres{E-Mail: gonzalo.olmo@uv.es. \vspace{-12pt}}

\abstract{We study the structure of a family of static, spherically symmetric space-times generated by an anisotropic fluid and governed by a particular type of $f(R)$ theory. We find that for a range of parameters with physical interest, such solutions represent black holes with the central singularity replaced by a finite size wormhole. We show that time-like geodesics and null geodesics with nonzero angular momentum never reach the wormhole throat due to an infinite potential barrier. For null radial geodesics, it takes an infinite affine time to reach the wormhole. This means that the resulting space-time is geodesically complete and, therefore, nonsingular despite the generic existence of curvature divergences at the \mbox{wormhole throat}.}


\begin{document}

\vspace{-12pt}
\section{Introduction}

Our understanding of the physical world is nowadays based on two basic pillars, namely, quantum theory and the notion of curved space-time. The  enormous individual success of these approaches contrasts with the difficulties that arise when one attempts to combine them. This is particularly evident when one explores quantum properties of black holes. Despite the classical robustness of these objects, which rapidly decay after they are formed into stationary configurations characterized by their mass, charge, and angular momentum \cite{Collapse1,Collapse2,Collapse3}, Hawking discovered that they are affected by a quantum instability due to the very existence of an event horizon \cite{Hawking}, with puzzling implications for the unitary evolution of quantum states \cite{Fabbri:2005mw}. This low energy phenomenon is sometimes regarded as the tip of a larger iceberg that dwells deeper down the black hole. The emitted Hawking quanta, which possess positive energy and cause a slow evaporation of the black hole, are correlated with negative energy modes that fall into the central singularity. The precise fate of the infalling quanta is completely unknown due to the breakdown of the geometric description as the singularity is approached. An improved understanding of the physical processes taking place in the innermost regions of black holes could thus help shed some new light on the evaporation process and the interaction between geometry and the quantum.

As a way to get a new perspective on the internal structure of black holes, we initiated a research program based on the notion of curved space-time but allowing the geometry to be slightly more general than the standard Riemannian framework of classical General Relativity. In this program we allow the affine connection to be determined by a variational principle rather than being fixed a priory by the Christoffel symbols of the space-time metric. This approach is supported, mainly, by the different mathematical and physical meanings that metric and affine connection have \cite{Zanelli,Olmo:2012yv}, and by the empirical evidence of the emergence of this type of geometries in condensed matter systems with microscopic defects \cite{LOR15}. If the space-time continuum that we perceive had a microscopic structure, a lesson that could be extracted from condensed matter systems is that metric-affine geometries could be the natural framework to formulate gravitational theories and their phenomenology \cite{Olmo:2011sw}.

By reconsidering the roles typically attributed to metric and affine structures in certain extensions of General Relativity (GR), we have been able to show that black hole singularities can be avoided in certain elementary configurations. To be more precise, we have considered electrovacuum space-times in theories of gravity including quadratic curvature terms \cite{Olmo:2013mla, Lobo:2013adx, Olmo:2011ja,or12a,or12b,or12c} as well as in others inspired by the Born-Infeld nonlinear theory of electrodynamics \cite{BI,ors13}. The novel and main result of our analysis is that space-time singularities can be avoided without necessarily removing curvature \mbox{divergences \cite{letter}}. The absence of singularities in these space-times is due to the fact that the GR point-like central singularity is generically replaced by a finite-size wormhole structure. We point out that such wormholes, as opposed to those typically considered in the framework of GR \cite{Wormholes}, do not violate any of the energy conditions. This mechanism to remove space-time singularities is in sharp contrast with the standard state-of-the-art of the field, where the widespread identification between curvature divergences and space-time singularities has developed multiple strategies focused on bounding curvature scalars as a way to obtain nonsingular configurations \cite{AB1,AB2,AB3,Ansoldi,Dymnikova}, a program that carries its own limitations, such as the difficulty to find theoretically well-grounded theories able to support such nonsingular solutions. To the light of our results, this identification has to be reconsidered, and further exploration of extensions of GR with independent metric and affine structures is needed.

In this work we show that even in a simpler scenario, such as in the well-known $f(R)$ theories of gravity (see \cite{or-review} for a review), one can find black hole configurations which represent nonsingular space-times. Classically, singular space-times are characterized by the existence of inextendible \mbox{paths \cite{Geroch:1968ut,Hawking:1973uf,Wald:1984rg}} (see also \cite{Curiel2009} and references therein). In particular, if physical observers (represented by time-like or null geodesics) start or terminate their paths at a finite value of their affine parameter, then the space-time is said to be singular. If those geodesics can be extended to arbitrarily large values of their affine parameter in the past and in the future, then the space-time is nonsingular.

Like in the more sophisticated models cited above, in the $f(R)$ case considered here the extendibility of geodesics is guaranteed by the existence of a robust wormhole geometry at the center of the black hole. As we will see, this wormhole is supported by an energy-momentum tensor with the form of an anisotropic fluid which satisfies all the energy conditions. Interestingly, we find that the resulting geometry near the wormhole is functionally identical (up to constant coefficients) to that found in configurations generated by an electric field governed by the Born-Infeld nonlinear theory of electrodynamics \cite{Olmo:2011ja}. This fact allows us to conclude that the wormhole structure is a  genuine and robust property of these theories, having very little sensitivity to the details of the particular matter source chosen. In the asymptotic region far away from the wormhole, however, the geometry is very dependent on the particular details of the matter source. As we shall show, this is due to the fact that it is possible to establish a direct correspondence between the anisotropic fluid used here and a certain family of nonlinear theories of electrodynamics, which reinforces the original motivation of \cite{Olmo:2011ja} to explore $f(R)$ theories coupled to that particular type of matter.

\section{Field Equations and Matter Source}

To fix ideas, the action we are interested in can be written as

\be \label{eq:action}
S=\frac{1}{2\kappa^2} \int d^4x \sqrt{-g} f(R) + S_m(g_{\mu\nu},\psi_m)
\en
where $\kappa^2$ is related to Newton's constant (in GR, $\kappa^2=8\pi G$), $g$ is the determinant of the space-time metric $g_{\mu\nu}$, $f(R)$ is a given function of the curvature scalar $R \equiv g^{\mu\nu} R_{\mu\nu}(\Gamma)$, and $S_m$ is the matter action depending on $g_{\mu\nu}$ and on the matter fields $\psi_m$. The key idea in our program is that we allow metric and affine connection $\Gamma \equiv \Gamma_{\mu\nu}^{\lambda}$ to be independent (Palatini or metric-affine approach), which means that both of them are governed by the variational principle. Using the fact that $\Gamma$ appears in the action \mbox{Equation (\ref{eq:action})} only through the Ricci tensor $R_{\mu\nu}(\Gamma)$, independent variations of (\ref{eq:action}) with respect to $g_{\mu\nu}$ and $\Gamma_{\mu\nu}^{\lambda}$ yield the field equations \cite{or-review}

\bea
f_R R_{\mu\nu}-\frac{f}{2} g_{\mu\nu}&=& \kappa^2 T_{\mu\nu} \label{eq:metric} \\
\nabla_{\lambda}(\sqrt{-g} f_R g^{\mu\nu})&=&0 \label{eq:connection}
\ena
where $T_{\mu\nu}=-\frac{2}{\sqrt{-g}} \delta S_m/\delta g_{\mu\nu}$ is the energy-momentum tensor of the matter. It is well known that the connection Equation (\ref{eq:connection}) can be formally solved by introducing a conformally related metric $h_{\mu\nu}=f_R g_{\mu\nu}$ such that this equation becomes $\nabla_{\lambda}(\sqrt{-h}  h^{\mu\nu})=0$, which implies that the connection is compatible with $h_{\mu\nu}$, \emph{i.e.}, that it can be written as the Christoffel symbols of $h_{\mu\nu}$, hence $R_{\mu\nu}(\Gamma)=R_{\mu\nu}(h)$. With this result, the remaining field Equation (\ref{eq:metric}) for Palatini $f(R)$ theories can be written in the following \mbox{compact form}:
\begin{equation} \label{eq:eom}
{R_\mu}^\nu (h)=\frac{1}{f_R^2}\left[\frac{f}{2}{\delta_\mu}^\nu+\kappa^2{T_\mu}^\nu \right]\
\end{equation}
Here we are denoting ${R_\mu}^\nu(h)=R_{\mu\alpha}(h)h^{\alpha\nu}$, with $R_{\mu\alpha}(h)$ the Ricci tensor of the metric $h_{\mu\nu}$.  One must bear in mind that tracing over Equation  (\ref{eq:metric}) with the metric $g^{\mu\nu}$ one gets $Rf_R-2f=\kappa^2 T$ (with $T$ the trace of the energy-momentum tensor), which is an algebraic relation leading to $R=R(T)$.  This implies that all the objects on the right-hand-side of Equation (\ref{eq:eom}) are functions of the matter. We also note that Equation (\ref{eq:eom}) represents a system of second-order equations for $h_{\mu\nu}$ (and also for $g_{\mu\nu}$ because they are algebraically related). The second-order character of the field equations is a robust property of Palatini theories \cite{or-review, Olmo:2009xy, Makarenko:2014lxa,Makarenko:2014nca, Odintsov:2014yaa, Jimenez:2015caa}.

Given that the conformal factor that relates $h_{\mu\nu}$ with $g_{\mu\nu}$ is a function of the trace $T$ of the matter sources, for traceless sources such as a standard Maxwell electromagnetic field the metric field equations boil down exactly to those of GR for arbitrary $f(R)$ theory. To avoid this degeneracy and explore new dynamics in black hole configurations, in \cite{Olmo:2011ja} we considered matter sources represented by nonlinear theories of electrodynamics, where the Maxwell action is replaced by a given function of the gauge field invariants. These theories yield, in general, a nonvanishing trace in four dimensions. Different new black hole configurations were found in that work, including one with unusual properties corresponding to the case of $f(R)=R-\lambda R^2$ coupled to the Born-Infeld theory of electrodynamics  \cite{BI}. At that time, we simply noticed that the central singularity had been shifted from the origin to a finite radius. With our subsequent analyses in other gravity theories, we can now revisit those results and interpret them under the light of a new family of black holes, namely, those with a finite-size wormhole replacing the point-like central singularity. We will see next that the same geometric properties found in the case of having a nonlinear electric field as matter source are also reproduced by an anisotropic fluid of the form
\begin{eqnarray}
{T_\mu}^\nu &=&\text{diag}[-\rho,P_r,P_\theta ,P_\varphi] \nonumber \\
&=& \text{diag}[-\rho,-\rho,\alpha \rho ,\alpha\rho]\  \label{eq:fluid}
\end{eqnarray}
with $\alpha$ a constant, being $\alpha=1$ equivalent to the case of a Maxwell electric field. To satisfy the energy conditions, one requires $0\leq \alpha \leq 1$. This fluid has recently been used in \cite{Shaikh:2015oha} to generate an interesting one-parameter family of wormholes in the framework of Born-Infeld gravity \cite{BIg1,BIg2}, and also considered in~\cite{hlms} to discuss generic conditions for generation of wormholes (let us also point out that such anisotropic fluids, which are believed to be relevant for the description of compact objects, can be modelled in terms of a perfect fluid, a standard electromagnetic field, and scalar field \cite{Visser2}, which could have relevant implications for the description of spherically symmetric space-times). We will see that something similar occurs in the case of $f(R)$ gravity, confirming in this way that the wormhole structure of the \mbox{$f(R)=R-\lambda R^2$} model (and possibly others as well) is rather insensitive to the details of the matter sourcing it, thus supporting its robustness.

\section{Structure Equations}

To proceed, we consider a line element for $h_{\mu\nu}$ of the form
\begin{equation}
d\tilde{s}^2=-A(x)e^{2\psi(x)}dt^2+\frac{1}{A(x)}dx^2+\tilde{r}^2(x) d\Omega^2 \  \label{eq:dst2a}
\end{equation}
where $\psi(x)$ and $A(x)$ are functions to be determined by solving the field Equation  (\ref{eq:eom}), which can be written as
\begin{equation}
{R_\mu}^\nu(h)=\frac{1}{f_R^2}\left(\begin{array}{cc} \left(\frac{f}{2}-\kappa^2\rho\right)\hat I & \hat 0 \\ \hat 0 & \left(\frac{f}{2}+\alpha\kappa^2\rho\right)\hat I \end{array}\right) \
\end{equation}
where $\hat I $ and $\hat 0$ are the identity and zero $2\times 2$ matrices, respectively. This structure of the field equations implies that ${R_t}^t-{R_x}^x=0$, which allows to set $\psi(x)\to 0$ and $\tilde{r}^2(x)\to x^2$ without loss of generality, turning the auxiliary line element Equation (\ref{eq:dst2a}) into a standard Schwarzschild-like form
\begin{equation}
d\tilde{s}^2=-A(x)dt^2+\frac{1}{A(x)}dx^2+x^2 d\Omega^2 \  \label{eq:dst2b}
\end{equation}

The line element for the space-time metric $g_{\mu\nu}$ can thus be written as
\begin{equation}
d{s}^2=\frac{1}{f_R}\left(-A(x)dt^2+\frac{1}{A(x)}dx^2\right)+r^2(x) d\Omega^2 \  \label{eq:ds2}
\end{equation}
where we have defined $r^2(x)=x^2/f_R$. Using this line element, the conservation equation of the fluid Equation (\ref{eq:fluid}) leads to $\rho(x)=C/r(x)^{2+2\alpha}$ \cite{Shaikh:2015oha}, where $C$ is a constant with dimensions (in the $\alpha=1$ case, $C$ would be proportional to the electric charge squared).  With this result, we find that for the model

\begin{equation}
f(R)=R-\lambda R^2
\end{equation}
we get $R=2(1-\alpha)\kappa^2\rho(x)$, and $f_R=1-(4\lambda)\kappa^2(1-\alpha)C/r(x)^{2+2\alpha}$, which can be written as \mbox{$f_R=1-1/z^{2+2\alpha}$} by defining $r=r_c z$ and $r_c^{2+2\alpha}\equiv (4\lambda)\kappa^2(1-\alpha)C$. Now, using the Ansatz $A(x)=1-2M(x)/x$, from the equation ${R_\theta}^\theta$ we get
\begin{equation}\label{eq:Mx}
\frac{8\lambda}{r_c^2}M_x=\frac{\left(\frac{1}{1-\alpha}-\frac{1}{2z^{2+2\alpha}}\right)}{z^{2\alpha}f_R} \
\end{equation}

It is now convenient to rescale $x$ as $x=r_c \tilde{x}$ to obtain

\begin{equation} \label{eq:xz}
\tilde{x}=z\sqrt{1-1/z^{2+2\alpha}}
\end{equation}
and $d\tilde{x}=[(1+\alpha/z^{2+2\alpha})/\sqrt{1-1/z^{2+2\alpha}}]dz$, and to write $M(x)$ as $M(x)=M_0+M_0\delta_1 G({x})$, with $\delta_1=r_c^3/(8\lambda M_0)$. With all this, Equation  (\ref{eq:Mx}) turns into
\begin{equation}\label{eq:Gz}
G_z=\frac{\left(1+\frac{\alpha}{z^{2+2\alpha}}\right)}{z^{2\alpha}f_R^{3/2}} \left[\frac{1}{1-\alpha}-\frac{1}{2z^{2+2\alpha}}\right]\
\end{equation}

This equation can be readily integrated, yielding
\begin{equation}\label{eq:G}
G_\alpha(z)=\frac{z^{-4 \alpha -1} \left(\frac{z^{2 \alpha +2} \sqrt{1-z^{-2 (\alpha +1)}} \left(2 \alpha ^2+\alpha +2 z^{2 \alpha +2}-3\right)}{z^{2 \alpha +2}-1}-\frac{8 \alpha  \left(\alpha ^2-1\right) \, _2F_1\left(\frac{1}{2},\frac{4 \alpha +1}{2 \alpha +2};\frac{6 \alpha +3}{2 \alpha +2};z^{-2 (\alpha +1)}\right)}{4 \alpha +1}\right)}{2 (\alpha -1) (2 \alpha -1)} \
\end{equation}
where $_2F_1\left(a, b, c; y\right)$ is a hypergeometric function. The above solution is valid for $\alpha\neq 1, 1/2$. As mentioned earlier, when $\alpha=1$, the fluid we are considering degenerates into an electric Maxwell field, thus yielding the Reissner-Nordstr\"{o}m solution of GR (with $f_R=1$, $r^2=x^2$, and $G(z)=-1/z$). For $\alpha=1/2$, the integration of $G_z$ must be performed separately and the result is
\begin{equation}
G_{\frac{1}{2}}(z)=\gamma-\frac{8 z^3-8 \sqrt{z^3-1} z^{3/2} \log \left(z^{3/2}+\sqrt{z^3-1}\right)+1}{6 \sqrt{1-\frac{1}{z^3}} z^3} \
\end{equation}
where $\gamma\equiv \frac{4}{3}-\frac{4 \log (2)}{3}$ is a constant necessary to match the GR solution in the region $z\gg 1$ (in GR $G_{\frac{1}{2}}(z)=2\ln z$).
For completeness, we mention that in GR we have $G^{GR}_\alpha(z)=\frac{z^{1-2\alpha}}{(1-\alpha)(1-2\alpha)}$, which is recovered from Equation (\ref{eq:G}) in the $z\gg1$ limit.

We note that the GR limit is recovered very quickly, just a few units away from $z=1$, which implies that the number and location of horizons in these solutions is almost coincident with those predicted by GR. In this sense, one can find configurations with two horizons, with one degenerate horizon, or without horizons, depending on the particular combination of parameters $(\alpha,\delta_1, \delta_2)$ chosen.

\section{Wormhole geometry.}

\begin{figure}[H]
\begin{center}
\includegraphics[width=0.5\textwidth]{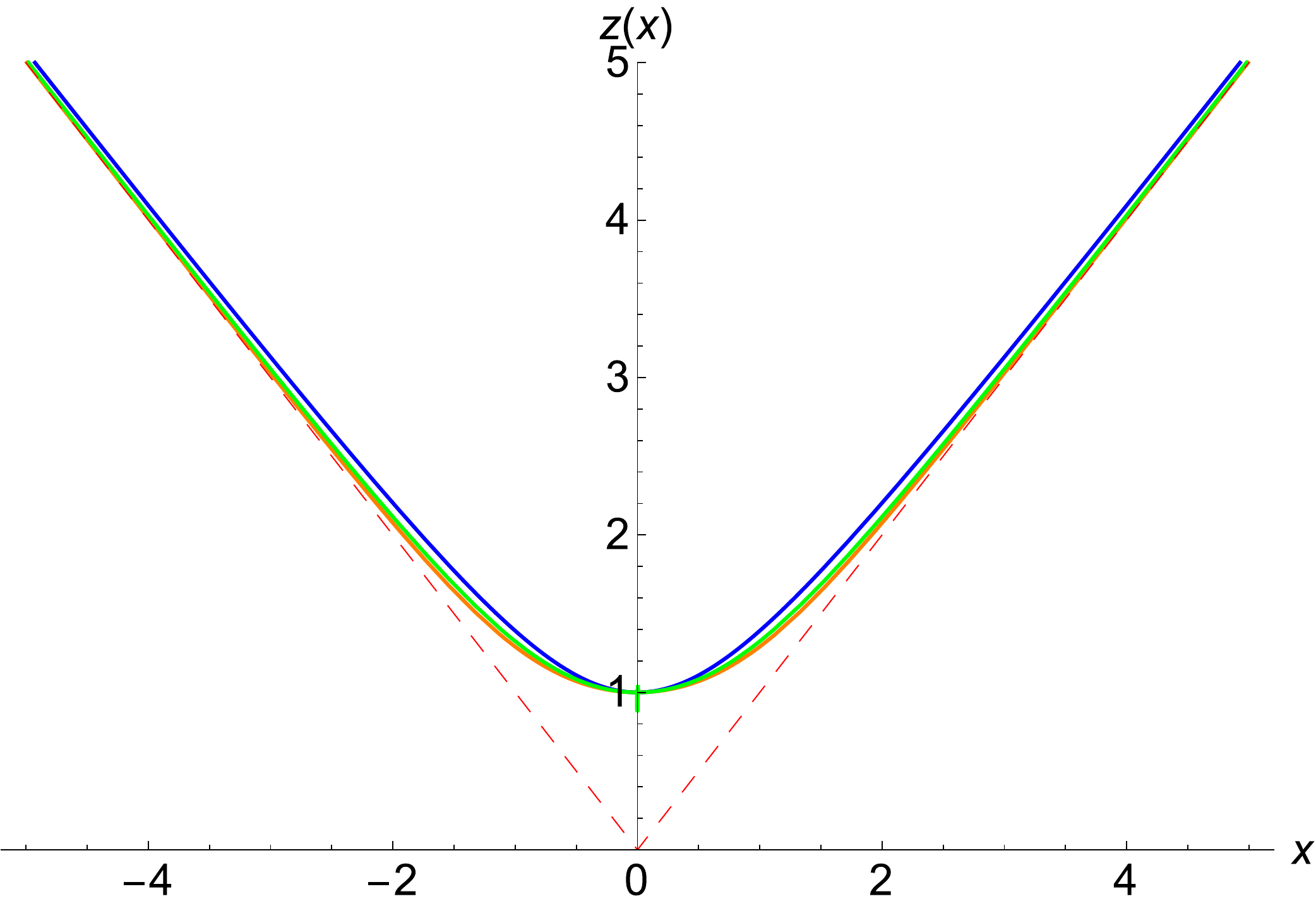}
\caption{Representation of the radial function $r(x)$ in units of the scale $r_c$ for the cases $\alpha=1/10, 1/2, 4/5$ (colors blue, green, and orange, respectively). The dashed lines represent the function $|x|$, corresponding to the General Relativity (GR) case, which is quickly recovered for $x >2 $. As can be seen, the dependence on $\alpha$ is very weak. }\label{fig:r(x)}
\end{center}
\end{figure}

As found in other Palatini theories \cite{or12a,or12b,or12c,ors13}, the existence of a wormhole can be inferred from the minimum of the area function $A=4\pi r^2(x)$ at $x=0$ (where $z=1$). The explicit relation $r=r(x)$ can be obtained for each $\alpha$ by inverting Equation (\ref{eq:xz}). For several values of $\alpha$ the result  is plotted in \mbox{Figure \ref{fig:r(x)}}. Near the wormhole throat, the metric component

\begin{equation}
g_{tt}=-\left(1-\frac{1+\delta_1 G(z)}{\delta_2 z f_R^{1/2}}\right)/f_R
\end{equation}
with $\delta_2=r_c/2M_0$ can be expanded as follows
\begin{eqnarray}
-g_{tt}&\approx& \frac{\delta_1}{ 8 \delta_2\left(1-\alpha^2 \right)}\frac{1}{(z-1)^2}-\frac{1-\frac{4 \sqrt{\pi } \alpha  (\alpha +1) \delta_1 \Gamma \left(\frac{6 \alpha +3}{2 \alpha +2}\right)}{\left(8 \alpha ^2-2 \alpha -1\right) \Gamma \left(\frac{5 \alpha +2}{2 \alpha +2}\right)}}{(2 \alpha +2)^{3/2} \delta_2}\frac{1}{(z-1)^{3/2}}+\frac{(3-6 \alpha ) \delta_1+4 (\alpha -1) \delta_2}{8 \left(\alpha ^2-1\right) \delta_2 }\frac{1}{(z-1)}\nonumber \\ &-&\frac{(6 \alpha +5) \left(\left(8 \alpha ^2-2 \alpha -1\right) \Gamma \left(\frac{5 \alpha +2}{2 \alpha +2}\right)-4 \sqrt{\pi } \alpha  (\alpha +1) \delta_1 \Gamma \left(\frac{6 \alpha +3}{2 \alpha +2}\right)\right)}{8 \sqrt{z-1} \left(\sqrt{2} (\alpha +1)^{3/2} (2 \alpha -1) (4 \alpha +1) \delta_2 \Gamma \left(\frac{5 \alpha +2}{2 \alpha +2}\right)\right)}\nonumber \\ &+&\frac{\left(-244 \alpha ^2+4 \alpha +171\right) \delta_1+64 \left(2 \alpha ^2+\alpha -3\right) \delta_2}{256 (\alpha -1) (\alpha +1) \delta_2}+O(\sqrt{z-1}) \
\end{eqnarray}

This expansion is functionally identical to that found in \cite{Olmo:2011ja} for this gravity model coupled to the nonlinear Born-Infeld theory of electrodynamics, with divergences of the form \mbox{$g_{tt}\approx a/(z-1)^2+b/(z-1)^{3/2}+\ldots$}, which induces similar divergences in the Kretschmann scalar, $K={R_\alpha}^{\beta\gamma\beta}{R^\alpha}_{\beta\gamma\beta}$, with leading order $K \simeq 1/(z-1)^{2}$.

The behavior far from the wormhole, however, does depend on the kind of matter source. In fact, in the Born-Infeld case of \cite{Olmo:2011ja}, the asymptotic geometry recovers the usual Reissner-Nordstr\"{o}m solution, $g_{tt}=-\left(1-\frac{1}{\delta_2 z}+\frac{\delta_1}{\delta_2z^2}\right)$ in the notation of this paper, whereas in the case of the fluid considered here, the asymptotic behavior is

\begin{equation} \label{eq:asymp}
g_{tt}=-\left(1-\frac{1}{\delta_2 z}+\frac{\delta_1}{\delta_2 (1-\alpha)(1-2\alpha)z^{2\alpha}}\right)
\end{equation}
which does depend on the specific fluid. This result can be understood as follows. Consider an electromagnetic field described by the action

\begin{equation}
S_m=\frac{1}{8\pi} \int d^4x \sqrt{-g} \varphi(X)
\end{equation}
where $\varphi(X)$ is a given function of the electromagnetic field invariant $X=-\frac{1}{2} F_{\mu\nu}F^{\mu\nu}$, where $F_{\mu\nu}=\partial_{\mu}A_{\nu}- \partial_{\nu}A_{\mu}$ is the field strength tensor of the vector potential $A_{\mu\nu}$. Different functions $\varphi(X)$ specify different models of nonlinear electrodynamics. For electrostatic configurations, $F^{tr} \equiv E(r)$, the components of the energy-momentum tensor become

\begin{equation}
{T_\mu}^{\nu}=\frac{1}{8\pi}\text{diag}[\varphi-2X\varphi_X, \varphi-2X\varphi_X, \varphi,\varphi]
\end{equation}

If we impose the equivalence of this energy-momentum tensor with that of the anisotropic fluid Equation  (\ref{eq:fluid}), one finds the following function

\begin{equation} \label{eq:Xalpha}
\varphi(X)=X^{p}
\end{equation}
where $p=(1+\alpha)/(2\alpha)$, as the matter nonlinear electromagnetic Lagrangian. This corresponds to a class of models falling within the general analysis carried out in \cite{dr1,dr2} for gravitating electromagnetic fields within GR. In addition, the properties of Equation (\ref{eq:Xalpha}) in different gravitational scenarios have been studied in detail in \cite{Hassaine1,Hassaine2,Hassaine3,Hassaine4,Hassaine5}. For our analysis, the relevant point here is that the asymptotic behavior of gravitational configurations sourced by the models Equation (\ref{eq:Xalpha}) is governed by the standard Schwarzschild mass term if \mbox{$1<p<3/2$}, corresponding in our case to $1/2<\alpha<1$ [see \mbox{Equation (\ref{eq:asymp})}], while those satisfying $p>3/2$ ($0<\alpha<1/2$) are asymptotically flat but not Schwarzschild-like, since they are governed by the charge term. In the case $p=1$ ($\alpha=1$) we get a standard Maxwell field and thus the standard asymptotic Reissner-Nordstr\"om behaviour of GR is recovered. This is the case of Born-Infeld electrodynamics mentioned above. Nevertheless, these asymptotic behaviors have no influence on the features of the wormhole structure, since this arises due to the interplay between the $f(R)$ gravity and the anisotropic fluid (nonlinear electromagnetic field) as the innermost region is approached for every $\alpha$, which indicates that the wormhole geometry is a robust structure insensitive to the matter source that generates it.

Let us point out that the existence of a wormhole structure seems to be a robust prediction of Palatini gravities, since they are also present in quadratic \cite{or12a,or12b,or12c,Olmo:2013mla,Lobo:2013adx} and Born-Infeld gravity theories \cite{ors13} sourced by (linear and non-linear) electromagnetic fields. The emergence of wormholes in scenarios with self-gravitating fields allows to interpret such solutions as geons in Wheeler's sense \cite{Wheeler:1955zz}, which further supports their topological nature. It should be noted that the quadratic $f(R)$ model considered here also exhibits nonsingular cosmological solutions with a bounce \cite{Barragan:2010qb,Barragan:2010uj,Barragan:2009sq} (see also \cite{Odintsov:2014yaa,Olmo:2014sra} for related results). Whether such bouncing solutions can be related to the interior of the black hole solutions obtained here is an issue that will be explored elsewhere.

\section{Geodesics}

The tangent vector for null and time-like geodesics ($k=0,1$ respectively) parameterized by an affine parameter $\lambda$ can be written using Equation (\ref{eq:ds2}) as \cite{geodesics,Wald:1984rg}
\begin{equation}
-k=-\frac{A(x)}{f_R}\left(\frac{dt}{d\lambda}\right)^2+\frac{1}{A(x)f_R}\left(\frac{dx}{d\lambda}\right)^2+r^2(x)\left(\frac{d\varphi}{d\lambda}\right)^2 \
\end{equation}
where the angular variable $\theta$ has been set to $\pi/2$ without loss of generality. Given the invariance of the metric under time translations and $\varphi$ rotations, we have two conserved quantities, namely, $E=\frac{A}{f_R}\frac{dt}{d\lambda}$ and $L=r^2\frac{d\varphi}{d\lambda}$. For time-like geodesics, the constants $E$ and $L$ represent the energy and angular momentum per unit mass, respectively. Using these conserved quantities, the relevant equation becomes
\begin{equation}\label{eq:dxdl2}
\left(\frac{dx}{d\lambda}\right)^2=E^2f_R^2-A(x)f_R \left(k+\frac{L^2}{r^2(x)}\right) \
\end{equation}

Instead of attempting to integrate this equation directly, it is convenient to replace the variable $x$ by $r=r_c z$ using the relation $\tilde{x}^2=z f_R^{\frac{1}{2}}$, which leads to an equation that relates $z$ and $\tilde \lambda=\lambda/r_c$:
\begin{equation}\label{eq:dzdl2}
\frac{d\tilde\lambda}{dz}=\pm \frac{f_R^{\frac{1}{2}}\left[1+\frac{zf_{R,z}}{2f_R}\right]}{\sqrt{E^2 f_R^2- A(z)f_R\left(k+\frac{L^2}{r_c^2z^2}\right)}} \
\end{equation}

From the positivity of the argument in the square root of the denominator of this equation, as well as from the right-hand side of Equation (\ref{eq:dxdl2}), it is rather obvious that for $k=1$ and/or $L\neq 0$, all such geodesics will bounce at a finite value $z>1$, which corresponds to the location at which $dx/d\lambda=0$ in Equation (\ref{eq:dxdl2}). Note, in this sense, that as $z\to 1$ the $E^2 f_R^2$ term goes to zero much faster than the other term in the square root. In fact, the leading order term of $A(z)f_R\left(k+\frac{L^2}{r_c^2z^2}\right)$ as $z\to 1$ is a positive constant, $\frac{(1+\alpha) \delta_1 }{(1-\alpha ) 2\delta_2}\left(k+\frac{L^2}{r_c^2}\right)$, which necessarily implies the vanishing of the square root at some $z>1$ for arbitrary $E$.  A similar behavior is also found in the standard Reissner-Nordstr\"{o}m solution of GR, where time-like geodesics and null geodesics with nonzero angular momentum cannot reach the central singularity due to an infinite potential barrier (see \cite{geodesics} and \cite{Chandra} for details). Null geodesics, however, do reach the central singularity in the Reissner-Nordstr\"{o}m case, and do it in a finite length of the affine parameter, implying in this way the existence of inextendible geodesics. As a result, the Reissner-Nordstr\"{o}m space-time is regarded \mbox{as singular}.

In our wormhole scenario, null radial geodesics satisfy the simpler equation
\begin{equation}\label{eq:dzdl2null}
\pm E\frac{d\tilde\lambda}{dz}= \frac{\left[1+\frac{zf_{R,z}}{2f_R}\right]}{f_R^{\frac{1}{2}}} \
\end{equation}
which admits an exact analytical solution of the form
\begin{equation}
\pm E\tilde\lambda (z)=-\frac{z}{\sqrt{1-z^{-2 (\alpha +1)}}}+2 z \, _2F_1\left(\frac{1}{2},-\frac{1}{2 (\alpha +1)};1-\frac{1}{2 (\alpha +1)};z^{-2 (\alpha +1)}\right) \
\end{equation}

\begin{figure}[H]
\begin{center}
\includegraphics[width=0.5\textwidth]{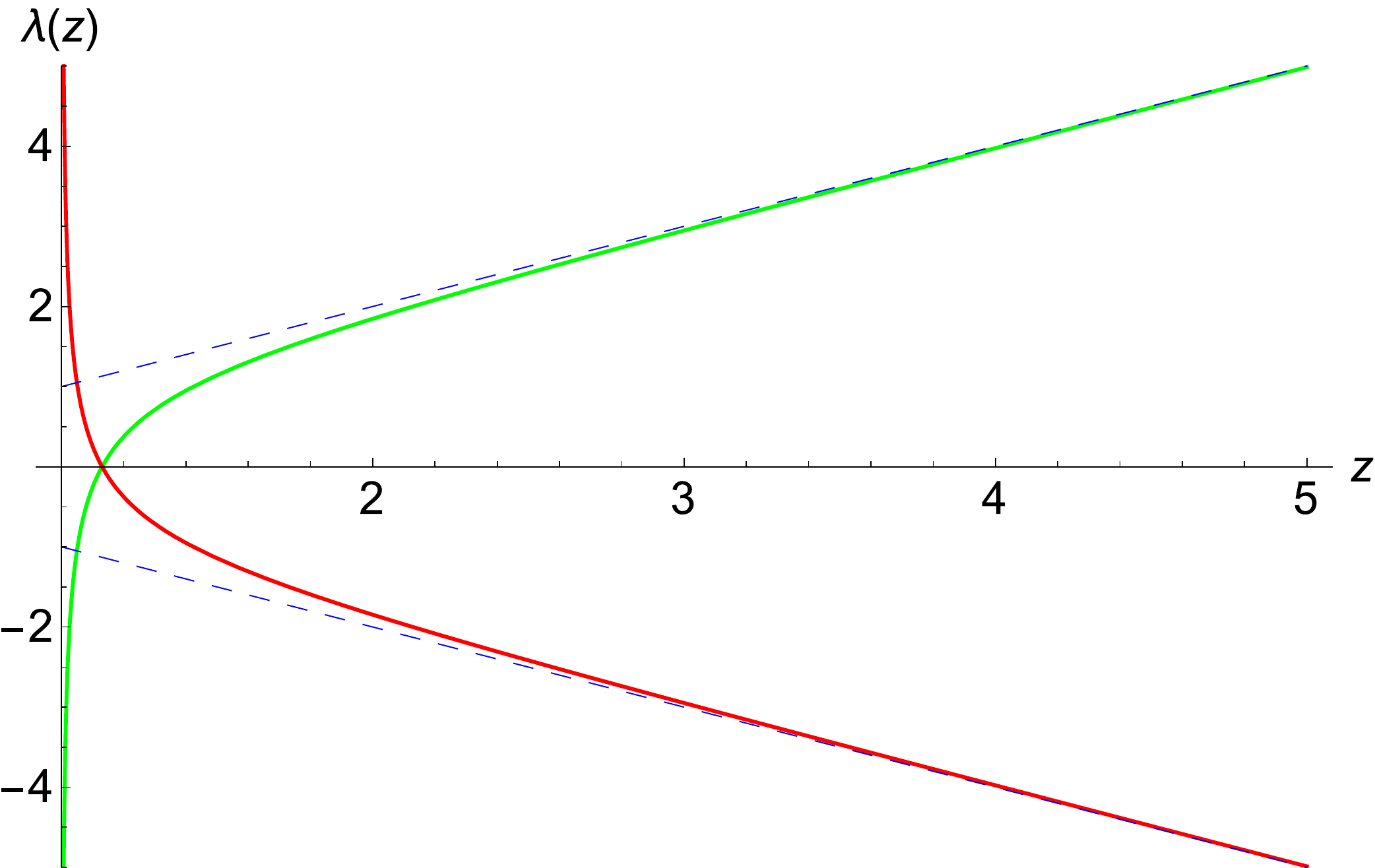}
\caption{Representation of outgoing (green) and ingoing (red) null radial geodesics with $E=1$ and $\alpha=4/5$. Similar results are found for any other value $0<\alpha<1$. }\label{fig:NullRadial}
\end{center}
\end{figure}

The asymptotic behaviors of this solution are the following: when $z\gg 1$, then $\pm E\tilde \lambda\approx z $, and when $z\to 1$ we find $\pm E\tilde \lambda\approx -\frac{1}{\sqrt{2 \alpha +2} \sqrt{z-1} }=-\frac{1}{|\tilde x|}$ (see Figure \ref{fig:NullRadial}). The divergence of the affine parameter on both limits, when $z\to \infty$ and when $z\to 1$, implies that null radial geodesics are complete. Note that in the limit $z\to 1$, light rays take an infinite (affine) time to reach the wormhole throat, where the curvature divergence is located. As a result, this space-time must be regarded as nonsingular, despite the existence of curvature divergences at the wormhole throat.

\section{Summary and Conclusions}

In this work, we have studied static, spherically symmetric geometries generated by a particular type of $f(R)$ theory of gravity formulated \`{a} la Palatini coupled to a one-parameter family of anisotropic fluids, which have also been put into correspondence with specific types of nonlinear theories of electrodynamics. We have shown that the resulting solutions represent wormholes and  that their geodesic structure is rather different from that discussed in \cite{letter, geodesics} for the Born-Infeld and Ricci-squared gravity models. In that case, null rays and time-like observers (in certain configurations) were able to go through the wormhole. In the case considered here, however, neither light rays nor time-like observers can ever reach the wormhole throat. In fact, we have shown that for null geodesics  the wormhole throat lies at an infinite affine distance. This point is very important because curvature scalars generically diverge at the throat. Given that this region lies beyond the reach of physical observers and light rays, the space-time must be regarded as nonsingular despite the existence of curvature divergences. This provides a different way to get rid of space-time singularities which complements the results presented in \cite{letter, geodesics} and confirms the versatility of Palatini theories of gravity to get rid of singularities in simple and clean scenarios. This is in sharp contrast with the difficulties found in the standard metric approach, where the higher-order character of the field equations largely prevents the construction of analytical solutions, unless additional constraints are imposed upon the theory and/or the solutions. Following the analysis carried out in \cite{letter} it would be interesting to explore the propagation of waves in this $f(R)$ scenario. Such study will be carried out elsewhere.

From our results, it follows that the emergence of the wormhole is a strongly non-perturbative phenomenon. In fact, a glance at  Figure \ref{fig:NullRadial} puts forward that the linear behavior $\lambda(z)\approx z$ of GR for null radial geodesics is valid in the $f(R)$ case from infinity down to the scale $z\approx 2$, with the wormhole located at $z=1$. This means that there are no local experiments that can be performed away from the wormhole that may be used to infer its existence, \emph{i.e.}, there is no way to distinguish between this $f(R)$ theory and  GR by looking at perturbative corrections. One must get right at the throat of the wormhole to perceive its existence. An important lesson we extract from this result is that GR can be trusted at high and very high energies because there are examples of theories  (like the one considered here) in which its shortcomings can be resolved by sudden non-perturbative effects.

\acknowledgments{Acknowledgments}
Gonzalo J. Olmo is supported by a Ramon y Cajal contract, the Spanish grant FIS2011-29813-C02-02, the Consolider Program CPANPHY-1205388, and the i-LINK0780 grant of the Spanish Research Council (CSIC). Diego Rubiera-Garcia is supported by the NSFC (Chinese agency) grants No. 11305038 and No. 11450110403, the Shanghai Municipal Education Commission grant for Innovative Programs No. 14ZZ001, the Thousand Young Talents Program, and Fudan University. Gonzalo J. Olmo and Diego Rubiera-Garcia also acknowledge support of CNPq project No. 301137/2014-5.


\authorcontributions{Author Contributions}

Gonzalo J. Olmo and  Diego Rubiera-Garcia conceived and wrote this paper. All authors have read and approved the final version.


\conflictofinterests{Conflicts of Interest}

The authors declare no conflict of interest.



\begin{thebibliography}{----}



\bibitem{Collapse1} Penrose, R. Gravitational collapse and space-time singularities. \emph{Phys. Rev. Lett.} \textbf{1965}, \emph{14}, 57-59.
\bibitem{Collapse2} Penrose, R. Gravitational collapse: The role of general relativity. \emph{Riv. Nuovo Cimento} \textbf{1969}, \emph{1}, 252-276.
\bibitem{Collapse3} Carter, B. Axisymmetric Black Hole Has Only Two Degrees of Freedom. \emph{Phys. Rev. Lett.} \textbf{1971}, \emph{26}, 331-333.


\bibitem{Hawking} Hawking, S. Black-hole evaporation.  \emph{Nature} \textbf{1974}, \emph{248}, 30-31.


\bibitem{Fabbri:2005mw}
  Fabbri, A.; Navarro-Salas, J. \emph{Modeling Black Hole Evaporation}; Imperial College Press: London, UK, 2005.


\bibitem{Zanelli}
Zanelli, J.
Lecture notes on Chern-Simons (super-) gravities. 2005,
arXiv:hep-th/0502193. arXiv.org e-Print archive. Available online: http://arxiv.org/abs/hep-th/0502193 (accessed on 22 Feb 2005).


\bibitem{Olmo:2012yv}
 Olmo,  G.J.
  Introduction to Palatini theories of gravity and nonsingular cosmologies.
 In \emph{Open Questions in Cosmology}; Gonzalo, J.O., Ed.; InTech Publishing: Rijeka, Croatia, 2012; ISBN 978-953-51-0880-1.

\bibitem{LOR15}  Lobo, F.S.N.; Olmo, G.J.; Rubiera-Garcia, D. Crystal clear lessons on the microstructure of spacetime and modified gravity.  \emph{Phys. Rev. D} \textbf{2015}, \emph{91},  124001.


\bibitem{Olmo:2011sw}
  Olmo, G.J.
  Palatini Actions and Quantum Gravity Phenomenology.
  \emph{JCAP} \textbf{2011}, \emph{1110}, 018.


\bibitem{or12a} Olmo, G.J.;  Rubiera-Garcia, D.
  Reissner-Nordstr\'om black holes in extended Palatini theories.
  \emph{Phys. Rev. D} \textbf{2012}, \emph{86}, 044014;
\bibitem{or12b} Olmo, G.J.; Rubiera-Garcia, D.
Nonsingular charged black holes \`{a} la Palatini.
\emph{Int. J. Mod. Phys. D} \textbf{2012}, \emph{21}, 1250067.
\bibitem{or12c} Olmo, G.J.; Rubiera-Garcia, D.
Nonsingular black holes in quadratic Palatini gravity.
\emph{Eur. Phys. J. C} \textbf{2012}, \emph{72}, 2098.


\bibitem{Olmo:2013mla}
  Olmo, G.J.; Rubiera-Garcia, D.
 Semiclassical geons at particle accelerators.
  \emph{JCAP} \textbf{2014}, \linebreak \emph{1402}, 010.


\bibitem{Lobo:2013adx}
  Lobo, F.S.N.; Olmo, G.J.; Rubiera-Garcia, D.
 Semiclassical geons as solitonic black hole remnants.
  \emph{JCAP} \textbf{2013}, \emph{1307}, 011.

\bibitem{Olmo:2011ja}
  Olmo, G.J.; Rubiera-Garcia, D.
  Palatini $f(R)$ Black Holes in Nonlinear Electrodynamics.
 \emph{ \mbox{Phys.  Rev.  D}} \textbf{2011}, \emph{84}, 124059.

\bibitem{BI} Born, M.; Infeld, L. Foundations of the new field theory. \emph{Proc. R. Soc. Lond. A} \textbf{1934}, \emph{144}, 425-451.

\bibitem{ors13}  Olmo, G.J.; Rubiera-Garcia, D.; Sanchis-Alepuz, H.
 Geonic black holes and remnants in Eddington-inspired Born-Infeld gravity.
  \emph{Eur.  Phys.  J.  C} \textbf{2014}, \emph{74}, 2804.

\bibitem{letter}  Olmo, G.J.; Rubiera-Garcia, D.; Sanchez-Puente, A.  Classical resolution of black hole singularities via wormholes. 2015, arXiv:1504.07015 [hep-th]. arXiv.org e-Print archive. Available online: http://arxiv.org/abs/1504.07015 (accessed on 27 Apr 2015).

\bibitem{Wormholes} Visser, M.  {\it Lorentzian Wormholes}; Springer-Verlarg: New York, NY, USA, 1996.

\bibitem{AB1} Ay\'on-Beato, E.; Garc\'ia, A. 	
Regular black hole in general relativity coupled to nonlinear electrodynamics. \emph{Phys. Rev. Lett.} \textbf{1998}, \emph{80}, 5056-5059.

\bibitem{AB2} Ay\'on-Beato, E.; Garc\'ia, A. 	
Nonsingular charged black hole solution for nonlinear source. \emph{Gen. Relativ. Gravit.} \textbf{1999}, \emph{31}, 629-633.

\bibitem{AB3} Ay\'on-Beato, E.; Garc\'ia, A.
New regular black hole solution from nonlinear electrodynamics. \emph{Phys. Lett. B} \textbf{1999}, \emph{464}, 25.

\bibitem{Ansoldi} Ansoldi, S.  Spherical black holes with regular center: A review of existing models including a recent realization with Gaussian sources. 2008, arXiv:0802.0330[gr-qc]. arXiv.org e-Print archive. Available online: http://arxiv.org/abs/0802.0330 (accessed on 4 Feb 2008).

\bibitem{Dymnikova} Dymnikova, I. Cosmological term as a source of mass. \emph{Class. Quant. Grav.} \textbf{2002}, \emph{19}, 725-740.

\bibitem{or-review} Olmo,  G.J. Palatini approach to modified gravity: f(R) theories and beyond.
\emph{Int. J. Mod. Phys. D} \textbf{2011}, \emph{20}, 413.


\bibitem{Geroch:1968ut}
  Geroch, R.P.
 What is a singularity in general relativity?
  \emph{Ann. Phys.}  \textbf{1968}, \emph{48}, 526.

  \bibitem{Hawking:1973uf}
  Hawking, S.W.;  Ellis, G.F.R. \emph{The Large Scale Structure of Space-Time}; Cambridge University Press:
Cambridge, UK, 1973.

\bibitem{Wald:1984rg}
  Wald, R.M. \emph{General Relativity}; University Press: Chicago, IL, USA,  1984.

\bibitem{Curiel2009}
Curiel, E.; Bokulich, P.  {\it Singularities and Black Holes}; Zalta, E.N., Ed.; The Stanford Encyclopedia of Philosophy: 2012. Available online: (http://plato.stanford.edu/archives/fall2012/\linebreak entries/spacetime-singularities/ (accessed on 29 Jan 2009),.

\bibitem{Makarenko:2014lxa}
  Makarenko, A.N.; Odintsov, S.D.; Olmo, G.J.
 Born-Infeld-$f(R)$ gravity.
  \emph{Phys.  Rev.  D} \textbf{2014},  \linebreak \emph{90}, 024066.

\bibitem{Makarenko:2014nca}
 Makarenko,  A.N.; Odintsov, S.D.; Olmo, G.J.
  Little Rip, $\Lambda$CDM and singular dark energy cosmology from Born-Infeld-$f(R)$ gravity.
  \emph{Phys. Lett. B} \textbf{2014}, \emph{734}, 36.

\bibitem{Odintsov:2014yaa}
  Odintsov, S.D.; Olmo, G.J.; Rubiera-Garcia, D.
  Born-Infeld gravity and its functional extensions.
  \emph{Phys.  Rev.  D} \textbf{2014}, \emph{90}, 044003.


\bibitem{Jimenez:2015caa}
  Jim\'{e}nez, J.B.; Heisenberg, L.; Olmo, G.J.
  Tensor perturbations in a general class of Palatini theories.
\emph{ JCAP} \textbf{2014}, \emph{1411}, 004.

\bibitem{Olmo:2009xy}
 Olmo, G.J.; Sanchis-Alepuz, H.; Tripathi, S.
 Dynamical Aspects of Generalized Palatini Theories of Gravity.
  \emph{Phys.  Rev. D} \textbf{2009}, \emph{80}, 024013.

\bibitem{Shaikh:2015oha}
  Shaikh, R.
  Lorentzian wormholes in Eddington-inspired Born-Infeld gravity. 2015,
  arXiv:1505.01314 [gr-qc]. arXiv.org e-Print archive.

\bibitem{BIg1} Deser, S.;  Gibbons, G.W.
Born-Infeld-Einstein actions?.  \emph{Class. Quantum Gravity} \textbf{1998}, \emph{15}, L35.

\bibitem{BIg2} Ba\~nados, M.;   Ferreira,  P.G.
Eddington's theory of gravity and its progeny. \emph{Phys. Rev. Lett.} \textbf{2010}, \emph{105}, 011101.


\bibitem{hlms} Harko, T.; Lobo, F.S.N.; Mak, M.K.; Sushkov, S.V.
  Wormhole geometries in Eddington-inspired Born-Infeld gravity. 2013,
  arXiv:1307.1883 [gr-qc].

\bibitem{Visser2} Boonserm, P.; Ngampitipan, T.; Visser, M. Modelling anisotropic fluid spheres in general relativity. 2015, arXiv:1501.07044[gr-qc].

\bibitem{dr1} Diaz-Alonso, J.; Rubiera-Garcia,  D. Electrostatic spherically symmetric configurations in gravitating nonlinear electrodynamics. \emph{Phys. Rev. D} \textbf{2010}, \emph{81},  064021.

\bibitem{dr2} Diaz-Alonso, J.; Rubiera-Garcia,  D. Asymptotically anomalous black hole configurations in gravitating nonlinear electrodynamics. \emph{Phys. Rev. D} \textbf{2010}, \emph{82},  085024.


\bibitem{Hassaine1}
Hassaine, M.; Martinez, C.
  Higher-dimensional black holes with a conformally invariant Maxwell source.
\emph{  Phys.  Rev. D} \textbf{ 2007}, \emph{75}, 027502.

\bibitem{Hassaine2}
 Hassaine, M.; Martinez, C.
  Higher-dimensional charged black holes solutions with a nonlinear electrodynamics source.
 \emph{ Class.  Quantum Gravity}   \textbf{2008}, \emph{25}, 195023.

\bibitem{Hassaine3}
Hendi, S.H.
 Topological black holes in Gauss-Bonnet gravity with conformally invariant Maxwell source.
  \emph{Phys.  Lett.  B} \textbf{2009}, \emph{677}, 123-132.

\bibitem{Hassaine4}
Hendi, S.H.; Rastegar-Sedehi, H.R.
  Ricci flat rotating black branes with a conformally invariant Maxwell source.
  \emph{Gen. Relativ. Gravit.}   \textbf{2009}, \emph{41}, 1355-1366.

\bibitem{Hassaine5}
Gonzalez, H.A.; Hassaine, M.; Martinez, C.
  Thermodynamics of charged black holes with a nonlinear electrodynamics source.
  \emph{Phys. Rev. D} \textbf{2009},  \emph{80}, 104008.


\bibitem{Wheeler:1955zz}
  Wheeler, J.A.
  Geons.
  \emph{Phys. Rev.}  \textbf{1955}, \emph{97}, 511.

\bibitem{Barragan:2010qb}
  Barragan, C.; Olmo, G.J.
 Isotropic and Anisotropic Bouncing Cosmologies in Palatini Gravity.
  \emph{Phys.  Rev. D} \textbf{2010},  \emph{82}, 084015.


\bibitem{Barragan:2010uj}
  Barragan, C.; Olmo, G.J.; Sanchis-Alepuz, H.
  Avoiding the Big Bang Singularity with Palatini f(R) Theories. 2010,
  arXiv:1002.3919 [gr-qc]. arXiv.org e-Print archive. Available online: http://arxiv.org/abs/1002.3919 (accessed on 20 Feb 2010).


\bibitem{Barragan:2009sq}
  Barragan, C.; Olmo, G.J.; Sanchis-Alepuz, H.
  Bouncing Cosmologies in Palatini f(R) Gravity.
 \emph{ Phys.  Rev.  D} \textbf{2009}, \emph{80}, 024016.


\bibitem{Olmo:2014sra}
 Olmo,  G.J.; Rubiera-Garcia, D.
  Brane-world and loop cosmology from a gravity-matter coupling perspective.
 \emph{ Phys.  Lett. B} \textbf{2015}, \emph{740}, 73.



 \bibitem{geodesics}  Olmo, G.J. Rubiera-Garcia, D.; Sanchez-Puente, A. Geodesic completeness in a wormhole space-time with horizons. In preparation, 2015.



\bibitem{Chandra}
Chandrasekhar, S. \emph{The Mathematical Theory of Black Holes}; Oxford University Press: New York, NY, USA, 1992.


\end{thebibliography}
\end{document}